**Title:** Consumers behavior of Portuguese wine

**Author:** Vítor João Pereira Domingues Martinho

Research Centre of the Polytechnic Institute of Viseu

Av. Cor. José Maria Vale de Andrade

Campus Politécnico

3504 - 510 Viseu

PORTUGAL

e-mail: vdmartinho@esav.ipv.pt

# Consumers behavior of Portuguese wine

**Abstract**

There are few papers about the consumption pattern of the Portuguese wine, using econometrics techniques. This work, pretend to analyze the consumers behavior of the wine produced in Portugal, determining the demand equation with panel data methods. There were used statistical data available in the Alentejo Regional Winegrowing Commission (CVRA) website. These data were obtained from a study about the market analysis, made, in 2009, by the A.C. Nielsen. The data are disaggregated by region and type of wine (Doc Verde+Regional Minho, Doc Região do Douro+Regional Terras Durienses, Doc Região da Bairrada+Regional Beiras, Doc Região do Dão+Regional Beiras, Doc da Região de Lisboa+Regional de Lisboa, Doc da Região do Tejo+Regional do Tejo, Doc da Região de Setúbal+Regional Terras do Sado, Doc Alentejo+Regional Alentejano e Doc da Região do Algarve+Regional do Algarve), year (2008 and 2009) and by form of consumption (take home, direct consumption and discount). This work found some linear regularity in the consumers behavior of the Portuguese wine.

## 1. Introduction

The wine production in Portugal represented, in 2007, a value obtained of about 13%, representing a significant part of the national agricultural sector and an increasing weight in this economic activity, from 9,1% in 1980-1986 to 15,4% in 2000-2007. The Portuguese vineyard occupies about 240 thousand hectares distributed by the different regions, occupying 6,9% of the agricultural surface and 2,6 of the continental area. In the agricultural year, 2008/2009, about 30 thousand farmers declared wine production. The contribution of the different regions to the national production of wine is diverse, representing the Douro 25%, Lisboa 17%, Beiras and Alentejo 14%, each one. Is significant the increase of the Portuguese wine certificate, corresponding to about 75% of the all wine production (IVV, 2009).

There are many factors which can influence the wine consumption, like the price, the quality, the income of the consumers, the price of the others products, etc. Sometimes is the place of origin that influences the option of the consumers, namely the region (Ribeiro and Santos, 2008), but also the country (Dahlström and Åsberg, 2009). The price is considered in some literature as the most important factor for the demand. In general, the French wines have more inelastic prices and the Portuguese, Spanish and Italian wines more elastic prices (Muhammad, 2011). The price of others products, as said before, can be, also, an important determinant of the demand, namely the price of the substitute wines, generating competition between brands and a partial loyalty in the market of wine (Torrisi et al., 2006). The wine quality, specifically when attested by recognized experts in good reviews, has a determinant contribution in increasing the demand (Friberg and Grönqvist, 2012).

Because there are few works about the role of the price in the consumption of the wine, in general (Lockshin and Corsi, 2012), and Portuguese wine, in particular, in this paper it is pretended to estimate the demand equation for the consumption of this agroindustrial product in Portugal, namely the relationship between the quantities consumed and the prices.

It was not easy find statistical data about the Portuguese wine consume and the associated prices, but was possible get some information in the Alentejo Regional Winegrowing Commission website about the different wine regions of Portugal, for different type of wine (Doc or Regional), for the years 2008 and 2009 and considering diverse forms of purchase.

The demand equation is well explained in the demand theory. This theory is part of a group of theories in the microeconomics explanations. The microeconomic theory explicates the behavior of the economic agents, with four theories, two to the consumers and two for the producers/sellers. The consumers behavior is explained by the demand and utility theories. The demand explicates that the consumers buy frequently goods and services to satisfy their daily needs, but the quantities bought of each one good and service depend of several factors, where the price play an important role. This relationship can be linear or not and is expected a negative correlation between the quantities bought and the associated prices. These negative relations are known as the demand law. There are some exceptions to the demand law, namely in the luxury and basic goods. In the luxury goods this happen, because high prices increase the demand. Purchase goods with high prices is a signal of status and great social position. In the basic goods, in particular the food goods, because despite the prices, the consumers have daily needs associated with these products. The utility theory adds the satisfaction and the consumer income as other factors. In other words, the consumers pretend to maximize the satisfaction with the consumption, but this is limited by their income. The two contributions for the producers/sellers are the supply and production theories. The supply theory explains that the producers/sellers are interested in to produce/sell more quantities when the prices are high. So, there are a positive relationship among the quantities produced/sold and the prices. The production theory says which the producers/sellers pretend to maximize the profits. In the maximum profit point the marginal income (variation of the production in function the variation of the production factors) is equal to the production factor price and the marginal income is decreasing. The interaction among the consumers and the producers/sellers define the market for each good or service. In practice is not only this interaction that define the market, but is this, more the national, European and international intervention with taxes and subsidies. There are different forms of market, but it is possible to define two groups of markets, the structures in perfect competition and the markets in imperfect competition. The markets in perfect competition do not exist in practice, only in theory to be a base to define and understand other markets. In imperfect competition there are many forms of markets, but the extreme are the markets in monopoly, where there is only one economic agent that dominates all the market.

**2. Data analysis**

The variables considered are, taking into account the objectives referred before, the quantities of wine bought (in volume, liters) and the prices (in euros). The first 10 figures and the results presented in the table 1 for the first 10 estimations, are obtained for ten levels of analyze, namely taking into account the type of wine (all certificate – Doc and Regional, only Doc or only Regional) and the different forms of purchase (discount – some supermarket; take home – buy and take to home from super, hypermarket and small shops; or direct consumption – restaurants, snacks). In each figure and in each estimation the data are disaggregated by year (2008 and 2009) and by wine region.

The figure 11 and the results obtained in the last estimation showed in the table 1 are found putting all the data, considered before for each one of the ten levels, together.

From the analysis of the data, considering the figures 1 to 11, is not easy to find some conclusions about the linear or not linear relationship between the quantities of wine bought and the respective prices. Is not easy, too, concluding about the negative correlation among these two variables. However, observing the figures 1, 2 the pattern is similar and seems to be possible find some signs of linear and negative relationship between the wine demand and the prices.

The figure 11 because, the number of observations, 174, shows more evident signs of some linearity and some negative relationship between the variables quantities and the prices, however the indications are not clear and evident.

Anyway, considering that the principal difference between the figure 1 and 2 is about the exclusion, in the figure 1, the purchases in the form of discount, the similar evolution of these two figures is signal of few important of the wine purchases in the form of price cut. So, for the consumers of the Portuguese wine the discounts have not a great importance.

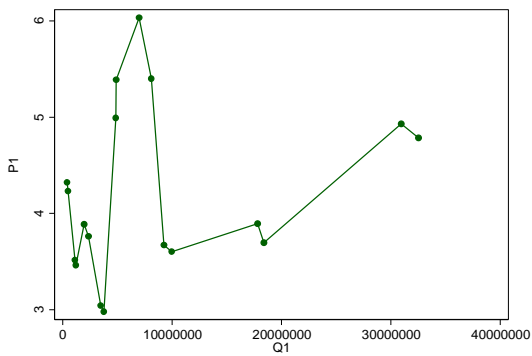

**Figure 1: Demand relation for all certificate wine and without discount**

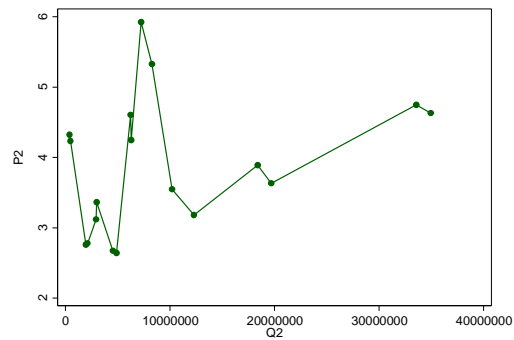

**Figure 2: Demand relation for all certificate wine and total aggregated**

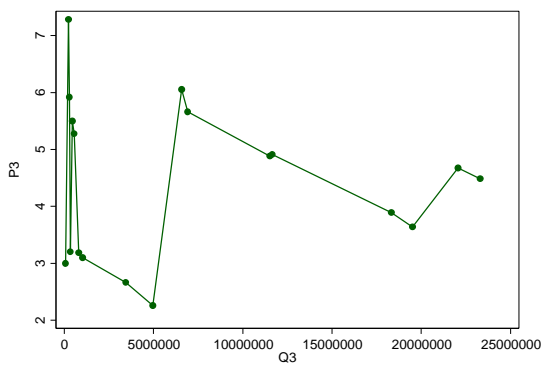

**Figure 3: Demand relation for Doc wine and total aggregated**

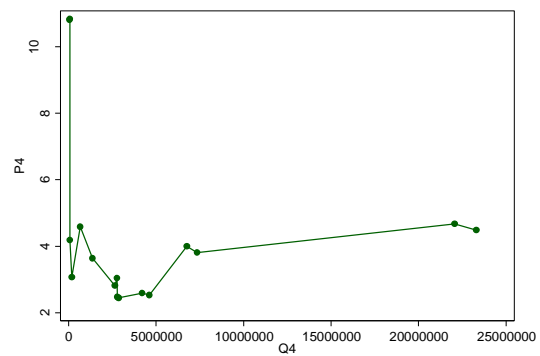

**Figure 4: Demand relation for Regional wine and total aggregated**

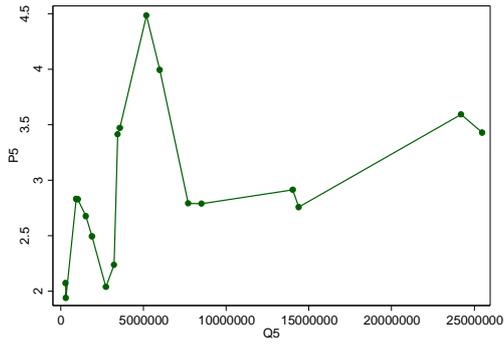

**Figure 5: Demand relation for all certificate wine and take home**

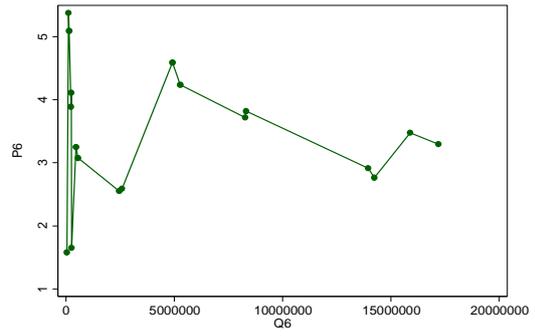

**Figure 6: Demand relation for Doc wine and take home**

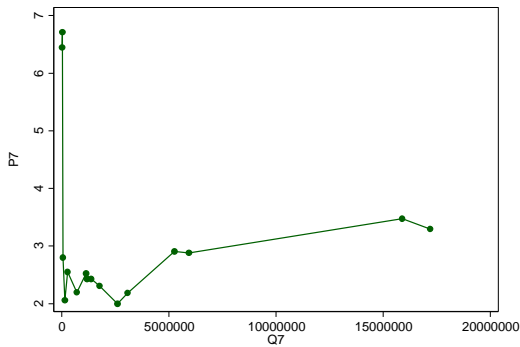

**Figure 7: Demand relation for Regional wine and take home**

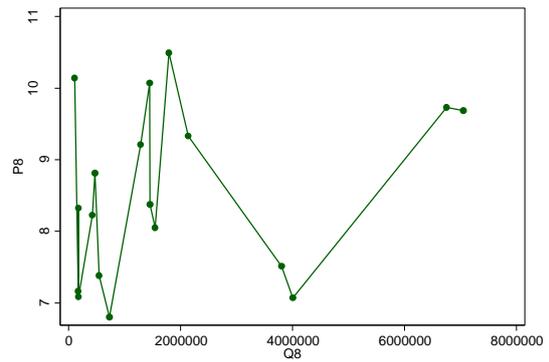

**Figure 8: Demand relation for all certificate wine and direct consumption**

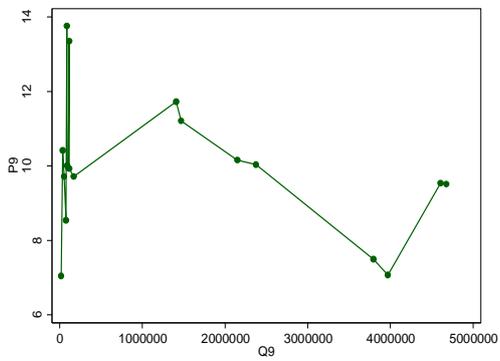

**Figure 9: Demand relation for Doc wine and direct consumption**

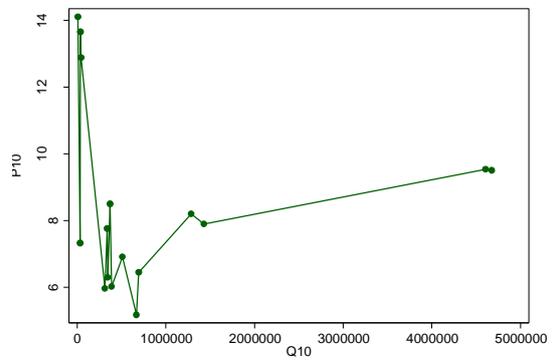

**Figure 10: Demand relation for Regional wine and direct consumption**

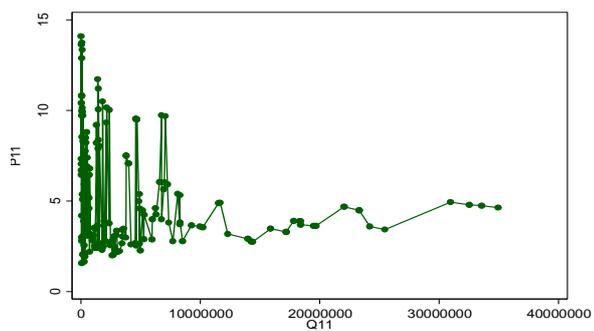

**Figure 11: Demand relation for total disaggregated for all different form of consume**

## 3. Estimations results

As referred before, for the data analysis, the results for the first and second estimations present some similarities and a similar pattern. Another time, considering that the differences between the statistical data for the first and second estimations are in the exclusion of the discount form of purchase, the results show the few relevance of the price cut for the consumers of Portuguese wine.

**Table 1: Results obtained with panel data, considering different regions and wine type, different forms of consumption and the years 2008 and 2009**

|  | Const.[1] | Coef.[2] | F/Wald(mod.)[3] | F(Fe_OLS)[4] | Corr(u_i)[5] | F(Re_OLS)[6] | Hausman[7] | $R^2$ [8] | N.O.[9] |
|---|---|---|---|---|---|---|---|---|---|
| **Without discount for all certificate wine** | | | | | | | | | |
| FE[10] | 16.400*[M] (5.380) | -1.814* [M] (-2.500) | 6.240* | 1137.410* | -0.390 | ------- | ------- | 0.438 | 18 |
| RE[11] | 15.800*[M] (3.400) | -1.675*[M] (-2.320) | 5.390* | ------- | ------- | 8.810* | 2.700 | 0.438 | 18 |
| **Total aggregated for all certificate wine** | | | | | | | | | |
| FE[10] | 19.600*[M] (5.91) | -2.512*[M] (-2.930) | 8.600* | 735.760* | -0.506 | ------- | ------- | 0.518 | 18 |
| RE[11] | 18.500*[M] (3.660) | -2.231*[M] (-2.560) | 6.540* | ------- | ------- | 8.650* | 2.930 | 0.518 | 18 |
| **Total aggregated for doc wine** | | | | | | | | | |
| FE[10] | 9.850*[M] (4.440) | -0.569[M] (-1.140) | 1.290 | 491.740* | -0.069 | ------- | ------- | 0.139 | 18 |
| RE[11] | 9.765*[M] (2.690) | -0.550[M] (-1.160) | 1.340 | ------- | ------- | 8.920* | 0.010 | 0.139 | 18 |
| **Total aggregated for regional wine** | | | | | | | | | |
| FE[10] | 7.313*[M] (4.650) | -0.504[M] (-1.410) | 1.980 | 757.350* | -0.071 | ------- | ------- | 0.220 | 16 |
| RE[11] | 7.233*[M] (2.340) | -0.486[M] (-1.480) | 2.180 | ------- | ------- | 7.960* | 0.020 | 0.220 | 16 |
| **Take home for all certificate wine** | | | | | | | | | |
| FE[10] | 11.700*[M] (4.490) | -1.629[M] (-1.840) | 3.370 | 724.000* | -0.498 | ------- | ------- | 0.297 | 18 |
| RE[11] | 10.800*[M] (2.860) | -1.337[M] (-1.490) | 2.210 | ------- | ------- | 8.670* | 3.950 | 0.297 | 18 |
| **Take home for doc wine** | | | | | | | | | |
| FE[10] | 9.205*[M] (3.640) | -1.138[M] (-1.550) | 2.400 | 833.020* | -0.115 | ------- | ------- | 0.231 | 18 |
| RE[11] | 8.975*[M] (2.810) | -1.071[M] (-1.580) | 2.500 | ------- | ------- | 8.950* | 0.060 | 0.231 | 18 |
| **Take home for regional wine** | | | | | | | | | |
| FE[10] | 5.145*[M] (2.550) | -0.522[M] (-0.800) | 0.640 | 344.300* | -0.122 | ------- | ------- | 0.083 | 16 |
| RE[11] | 4.925[M] (1.810) | -0.451[M] (-0.770) | 0.600 | ------- | ------- | 7.900* | 0.060 | 0.083 | 16 |
| **Direct consumption for all certificate wine** | | | | | | | | | |
| FE[10] | 2.683*[M] (4.340) | -0.093[M] (-1.280) | 1.630 | 540.920* | -0.316 | ------- | ------- | 0.169 | 18 |
| RE[11] | 2.622*[M] (2.700) | -0.085[M] (-1.190) | 1.410 | ------- | ------- | 8.420* | 0.570 | 0.169 | 18 |
| **Direct consumption for doc wine** | | | | | | | | | |
| FE[10] | 1.646*[M] (2.770) | -0.024[M] (-0.400) | 0.160 | 932.160* | 0.334 | ------- | ------- | 0.130 | 18 |
| RE[11] | 1.741*[M] (2.080) | -0.034[M] (-0.570) | 0.330 | ------- | ------- | 8.820* | 0.710 | 0.130 | 18 |
| **Direct consumption for regional wine** | | | | | | | | | |
| FE[10] | 1.074*[M] (6.240) | -0.011[M] (-0.530) | 0.280 | 487.990* | -0.040 | ------- | ------- | 0.039 | 16 |
| RE[11] | 1.073[M] (1.820) | -0.010[M] (-0.540) | 0.290 | ------- | ------- | 7.920* | 0.000 | 0.039 | 16 |
| **Total disaggregated for all different form of consume** | | | | | | | | | |
| FE[10] | 5.848*[M] (14.990) | -0.126[M] (-1.720) | 2.970 | 581.180* | 0.192 | ------- | ------- | 0.059 | 174 |
| RE[11] | 6.030*[M] (7.010) | -0.160*[M] (-2.260) | 5.090* | ------- | ------- | 86.110* | 3.700 | 0.059 | 174 |

Note: 1, Constant; 2, Coefficient; 3, Test F for fixed effects model and test Wald for random effects; 4, Test F for fixed effects or OLS (Ho is OLS); 5, Correlation between errors and regressors in fixed effects; 6, Test F for random effects or OLS (Ho is OLS); 7, Hausman test (Ho is GLS); 8, R square; 9, Number of observations; 10, Fixed effects model; 11, Random effects model; *, Statistically significant at 5%; [M] Values in million.

All the others forms of buy Portuguese wine do not present statistical significance for the coefficient of estimation, sign of no linearity, or lack of observations or few importance of the price in the justification of the wine consumption in Portugal for these forms of purchase. This is consistent with the first conclusion of not very importance of the prices in the explanation of the wine spending.

However in the last estimation, with 174 observations, it was found signs of the demand explanation by the respective prices, but the evidences, despite statistically significant, are not very strong.

On other side the values of the $R^2$ are not highs, what is reflex of little degree of explanation of the wine demand models considered. In this line, the values for the coefficients of the constant part are very high and with statistical significance, signs of lack of variables. Despite this results, the statistical test show more importance of the random effects than fixed effects.

Anyway, They are need more variables to explain the wine consumption in Portugal, because the prices have some importance in any circumstances, but in a disaggregated level there will be others factors which will be able to explicate the consume of Portuguese wine. In this line, is important to say that in every estimations the statistical tests reject the hypotheses of not fixed or random effects.

In the table 2 is possible to observe a first approach to the examination of the effects from the region of origin in the consumption of Portuguese wine, using variables dummies for each Continental wine region (D1 – Minho, D2 – Douro, D3 – Bairrada, D4 – Dão, D5 – Lisboa, D6 – Tejo, D7 – Setúbal, D8 – Alentejo, D9 – Algarve). From the results is possible to conclude about the negative effects in the consumption represented by the dummies 3 and 9, and the positive effect of the name Alentejo (dummy 8). Anyway, as said before, this is a first approach and it will be need more research about this issue.

**Table 2: Results obtained with panel data, all the data disaggregated, for the years 2008 and 2009 and with variables dummies**

| | Const.[1] | Coef.[2] | F/Wald(mod.)[3] | $D_1$[4] | $D_2$[4] | $D_3$[4] | $D_4$[4] | $D_5$[4] | $D_6$[4] | $D_7$[4] | $D_8$[4] | $D_9$[4] | $R^2$[5] |
|---|---|---|---|---|---|---|---|---|---|---|---|---|---|
| Total disaggregated for all different form of consume with variables dummies |||||||||||||||
| RE[6] | 9.262*[M] (2.190) | -0.195*[M] (-2.790) | 84.430 | 0.920[M] (0.200) | -4.290[M] (-0.950) | -7.522*[M] (-2.040) | 0.414[M] (0.110) | -6.241[M] (-1.380) | -6.983[M] (-1.540) | -3.195[M] (-0.700) | 8.265**[M] (1.820) | -7.847**[M] (-1.730) | 0.510 |

Note: 1, Constant; 2, Coefficient; 3, Test Wald for random effects; 4, Variables dummies for each wine region; 5, R square; 6, Random effects model; *, Statistically significant at 5%; **, Statistically significant at 10%; [M], Values in million.

## 4. Conclusions

The wine production in Portugal is a strategic cluster for the national economy and for the agricultural sector. All the contribution for improve the knowledge about the Portuguese wine sector are welcome. The works about the wine economy and policy are not easy, because the lack of statistical data, namely those available for the public in general. However, there is, already, some information which must be complemented in the future with other information and put available to the researchers with gains for everybody, in particular for the sector.

Despite the difficulty in find statistical information it was possible obtain some important conclusions from the data analysis and from the results get with the estimations. Nevertheless, will be important build other variables for the quality, consumers income, experts opinions, spatial autocorrelation and the influence of the origin wine region, and analyze the relationship between these variables and the demand for the Portuguese wine.

It was possible conclude about the few importance of the low prices (discounts) in the wine consumption in Portugal and about the relative few importance of the prices in the explanation of the Portuguese wine demand, taking into account the values of the $R^2$. On the other hand the wine demand is better explained by random effects than by fixed effects, but in every situation the demand model OLS (ordinary last square) was rejected by the statistical tests.

To consolidate these conclusions is important find in the future data for more years, not only to increase the number of observations, but also to get more regularity in the statistical information. Anyway, the conclusions obtained here furnish important contributions for the next research works, because give some orientations, namely about the significance of the prices in the explanations of the demand of Portuguese wines.

Considering the wine production an agroindustrial sector and considering the Portuguese rural context in the interior of Portugal, the contribution of the wine sector can be a crucial help for the balanced and sustainable development, between the urban and the rural reality and among the littoral and the interior contexts. Is important find strategies to the interior and rural development of the Portuguese economy, where the umbrella territorial brands and the territorial marketing are determinant. In this line, the wine, because it characteristics and it consumers, will be decisive for this balanced social, economic and environment growth and development. The wine is today a fashion product and is easily related with the tourism, industry, heritage, handicraft and environment. In the Douro region, for example, there are many examples of relation between the wine production and others activities and presently the Douro region is heritage of humanity. Another example are the wine routes existent in almost every wine regions. Other example are the small shops, with wine bottles, in some known touristic places, where the tourists must to pass to go out. These activities are crucial to create employment and put people in the more unfavorable zones. Creating work and putting people in the rural areas have the advantage of avoid the desertification and prevent some natural disasters, as the fires. An important part of the national and European subsidies to the farmers income is because this contribution of the agricultural producers to the sustainable development of unfavorable territories.

**References**

CVRA (2009). *Vendas dos Vinhos do Alentejo no Retalho e Restauração*. Source: A.C.Nielsen. Consulted in 27/12/2012 in the following website: http://www.vinhosdoalentejo.pt/media/documents/20120822_f76f6e.pdf.

Dahlström, T. and Åsberg, E. (2009). *Determinants of Demand for Wine – price sensitivity and perceived quality in a monopoly setting.* CESIS Electronic Working Paper Series, Paper No. 182.

Friberg, R. and Grönqvist, E. (2012). *Do Expert Reviews Affect the Demand for Wine?*. American Economic Journal: Applied Economics, 4, 1, 193-211.

IVV, IP (2009). *A Produção de Vinho em Portugal*. Factos e Números, Nº1, MADRP.

Lockshin, L. and Corsi, A.M. (2012). *Consumer behaviour for wine 2.0: A review since 2003 and future directions.* Wine Economics and Policy, In Press, Accepted Manuscript.

Muhammad, A. (2011). *Wine demand in the United Kingdom and new world structural change: a source-disaggregated analysis*. Agribusiness, 27, 1, 82–98.

Ribeiro, J.C. and Santos, J.F. (2008). *Portuguese quality wine and the region-of-origin effect: consumers´ and retailers´ perceptions.* NIPE WP 11/2008.

Torrisi, F.; Stefani, G.; and Seghieri, C. (2006). *Use of scanner data to analyze the table wine demand in the Italian major retailing trade*. Agribusiness, 22, 3, 391–403.